\documentclass[twocolumn,showpacs,aps]{revtex4}
\usepackage{epsf}
\usepackage{dcolumn}
\begin{document}
\title{Growth and Structure of Stochastic Sequences}
\author{E.~Ben-Naim}
\affiliation{Theoretical Division and Center for Nonlinear Studies,
Los Alamos National Laboratory, Los Alamos, New Mexico 87545}
\author{P.~L.~Krapivsky}
\affiliation{Center for Polymer Studies and Department of Physics,
Boston University, Boston, Massachusetts 02215} 
\begin{abstract}
 We introduce a class of stochastic integer sequences.  In these
 sequences, every element is a sum of two previous elements, at least
 one of which is chosen randomly.  The interplay between randomness
 and memory underlying these sequences leads to a wide variety of
 behaviors ranging from stretched exponential to log-normal to
 algebraic growth.  Interestingly, the set of all possible sequence
 values has an intricate structure.  \\
\end{abstract}
\maketitle
 
Integer sequences underly many problems in combinatorics, computer
science, and physics, with new beautiful sequences continuing to
emerge \cite{njas}. Sequences are typically deterministic.  Meanwhile,
stochastic sequences are just as ubiquitous, occurring in random
processes such as the random walk.  Stochastic sequences usually arise
in very different contexts, and hence are rarely compared with their
deterministic counterparts.  In this article, we demonstrate how rich
such a comparison can be.

Consider the Fibonacci numbers, $F_n=F_{n-1}+F_{n-2}$, that describe
for example the number of leaves in plants and the number of ancestors
of a drone \cite{conway,http,rad}.  As every element depends on the
previous two, a natural stochastic generalization is $x_n=x_{n-1} \pm
x_{n-2}$, where addition and subtraction are chosen with equal
probabilities \cite{F,ET,V,SK} (similar sequences also describe
one-dimensional disordered systems \cite{l,cpv}). The resulting
sequences are intriguing. While the sequences still grow
exponentially, the ratio $x_n/x_{n-1}$ approaches a stationary
distribution that possesses singularities at all rational values
\cite{V,SK}.

Inspired by this richness, we consider an alternative form of
stochasticity, namely, one that does {\em not} require subtraction and
therefore more similar in spirit to the original deterministic
sequence.  Relaxing the rule that every element depends only on the 
preceding two elements, we arrive at the following additive stochastic 
rules
\begin{eqnarray}
\label{rule}
x_n=\cases{x_{n-1}+x_q &(model I);\cr 
           x_p+x_q     &(model II).}
\end{eqnarray}
In model I, we take the preceding element $x_{n-1}$ and another one
whose index $q$ is randomly chosen between $0$ and $n-1$.  In model
II, both indices $p$ and $q$ are chosen randomly, $0\leq p,q\leq n-1$. 
Without loss of generality, the first element is set to unity, 
$x_0=1$.  Consequently $x_1=2$, while the next elements are
stochastic.  The number of possible sequences increases as $n!$ and
$n!^2$ for models I and II, respectively.  Rule I leads to
monotonically increasing sequences; sequences generated by rule II
increase only on average.

The most basic characteristic, the average of the $n$th element,
$A_n=\langle x_n\rangle$, can be determined analytically.  For model I,
it satisfies the linear recursion relation
\begin{equation}
\label{avI-rec}
A_n=A_{n-1}+{1\over n}\sum_{j=0}^{n-1}A_j.
\end{equation}
Comparing this with the recursion for $A_{n+1}$, we eliminate the
summation and obtain a Fibonacci-like recursion relation with
$n$-dependent coefficients
\hbox{$A_{n+1}-2A_n+A_{n-1}=(n+1)^{-1}A_{n-1}$}. We are primarily
interested in the large-$n$ behavior, and hence, we treat $A$ and $n$
as continuous variables. The above difference equation reduces to the
differential equation $A''=(A-A')/n$ (here $'\equiv d/dn$). Using the
WKB method \cite{bo}, we obtain the $n$-dependence 
\begin{equation}
\label{av-I}
A_n\simeq an^{-1/4}\,\exp\left(2\sqrt{n}\,\right).
\end{equation}
The amplitude $a\approx 0.1711$ is determined numerically. We see 
that the long range memory leads to considerably slower stretched
exponential growth compared with the exponentially growing Fibonacci
numbers.

\begin{figure}
\centerline{\epsfxsize=7.6cm\epsfbox{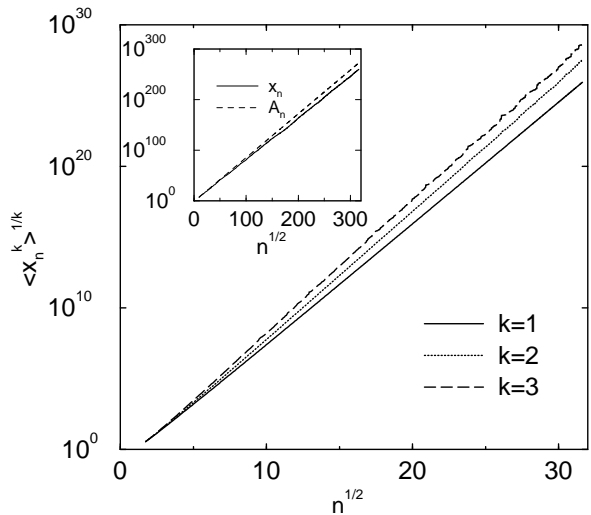}}
\caption{The moments $\langle x_n^k\rangle^{1/k}$ versus $n$ for model
I. The moments were obtained from an average over $10^8$
realizations. The inset compares the growth of an individual
realization $x_n$ with the average $A_n$.}
\end{figure}

Does the average characterize the growth of an actual sequence? If
yes, this would imply that the normalized moments $\langle
x_n^k\rangle/\langle x_n\rangle^k$ approach finite values
asymptotically.  Figure 1 shows otherwise: the higher order moments
grow according to
\begin{equation}
\label{M-grow}
\langle x_n^k\rangle\propto \exp\left(\beta_k \sqrt{n}\right),
\end{equation}
with $\beta_k>k \beta_1$ (for the lowest moments, we find $\beta_k=2$,
$4.3$, and $6.5$).  This so-called ``multiscaling'' indicates that a
typical sequence may greatly depart from the average. Therefore, the
average (\ref{av-I}) is insufficient to describe a typical
sequence. These results are similar in spirit to the behavior of the
random Fibonacci sequences where the typical growth is $x\propto
\exp(\gamma n)$, with nontrivial Lyapunov exponent $\gamma$, while
$\langle x_n^k\rangle\propto \exp(\gamma_k n)$ with $\gamma_k\ne
k\gamma_1$.

Interestingly, an individual realization grows slower than the
average (see the inset to Fig.~1) 
\begin{equation}
\label{typ-grow}
x_n\propto \exp\left(\beta\sqrt{n}\,\right),
\end{equation}
with the Lyapunov exponent $\beta\approx 1.889$.  This coefficient was
determined by studying the variable $\ln x_n$. As shown in Fig.~2,
this variable is Gaussian distributed
\begin{equation}
\label{phiz-I}
P_n(\ln x_n)\propto \exp\left[-{\left(\ln x_n-b_n\right)^2\over 2\Delta^2_n}\right].
\end{equation}
The average and the variance of $\ln x_n$ grow with $n$ according to
$b_n\simeq \beta n^{1/2}$ and $\Delta^2_n\simeq \sigma^2 n^{1/2}\propto
b_n$, respectively.  Eventually, the random variable $y=\ln x_n/n^{1/2}$ 
becomes deterministic, $y\to \beta$ as $n\to\infty$.
Similar behavior, including the Gaussian fluctuations and
the relation between the variance and the average, is also found in
one-dimensional localization problems \cite{o,ataf}.

\begin{figure}
\centerline{\epsfxsize=7.6cm\epsfbox{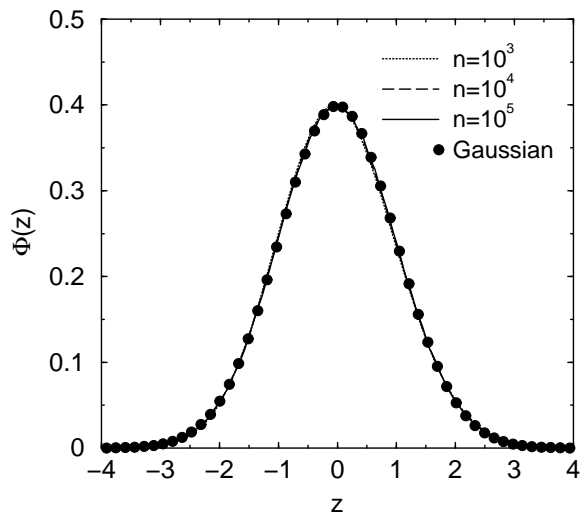}}
\caption{The scaling distribution underlying $P_n(\ln x_n)$. Shown is
the scaling function $\Phi(z)$ versus the scaling variable $z=(\ln
x_n-b_n)/\Delta_n$.  The distributions were calculated from $10^7$
realizations.  A Gaussian is also shown for
reference.}
\end{figure}

One can calculate the probability distribution $P_n(x)$ for extremal
values of $x_n$.  The minimal value $n+1$ is obtained by choosing
$q=0$ at every step $k=1,\ldots,n$.  Similarly, the maximal value
$2^n$ is obtained by choosing $q=k-1$ at every step.  Hence,
$P_n(n+1)=P_n(2^n)=1/n!$. Further extremal cases can be evaluated
manually, and for example, $P_n(n+2)=P_n\left(3\cdot
2^{n-2}\right)={n-1\over n!}$. However, these extremal probabilities
do not elucidate the typical behavior (\ref{typ-grow}).

It is interesting to study the set of all possible sequence values.
Let $\Omega_n$ be the support of the probability distribution
$P_n(x)$.  For small $n$, we have $\Omega_0=\{1\}$, $\Omega_1=\{2\}$,
and furthermore, 
\begin{eqnarray}
\label{Omega} 
\Omega_2&=&\{3,4\}\nonumber\\
\Omega_3&=&\{4,5,6,*,8\}\nonumber\\
\Omega_4&=&\{5,\ldots,10,*,12,*,*,*,16\}\nonumber\\
\Omega_5&=&\{6,\ldots,18,*,20,*,*,*,24,*,\ldots,*,32\}.
\end{eqnarray}
Determination of the sets $\Omega_n$ requires enumeration of all $n!$
histories, and we computed them up to $n=15$.  The simplest feature is
the set size $\Gamma_n$, listed in table I.  The set $\Omega_n$ always
begins with a subsequence of $B_n$ consecutive integers. For $n\geq
3$, the sets $\Omega_n$ have gaps, i.e., strings of missing elements
denoted by $*$ in Eq.~(\ref{Omega}).  The number of gaps $G_n$ is
listed in table I. The three sequences $\Gamma_n$, $B_n$, and $G_n$,
all grow exponentially with $n$. For example, $\Gamma_n\propto
\lambda^n$ with $\lambda\approx 1.78$.  We conclude that the sets
$\Omega_n$ contain a number of nontrivial deterministic integer
sequences including $\Gamma_n$, $B_n$, and $G_n$.

Remarkably, the sets $\Omega_n$ have an intricate structure.  For
instance, for $n=8$ the sequence of the gap lengths contains $G_8=18$
elements as follows
\begin{eqnarray}
\label{G8} 
\{1,1,1,1,1,3,1,1,3,3,7,7,1,3,7,15,31,63\}.  
\end{eqnarray}
Generally, going in reverse direction (from $2^n$ to $n+B_n$) one
observes a family of consecutive gaps of lengths $2^{n-2}-1,
2^{n-3}-1,\ldots,1$ separated by single elements.  Then, there is a
$3$-element sequence, followed by a second family of {\it twin} gaps,
$2^{n-5}-1,2^{n-5}-1,\ldots,1,1$.  All these gaps are separated by
single elements.  Next, there is a $5$-sequence, followed by a family
of {\it triplet} gaps, again separated by single elements [this family
has not yet formed for $n=8$, Eq.~(\ref{G8})].  There is also a fourth
family of {\it quadruplet} gaps with an intertwined pattern.  The
complexity of this gap-sequence structure increases rapidly, and
eventually, gaps of even length appear.

Naively, one may probe the probability distribution via a mean-field
description that ignores the sequence history altogether.  In this
approximation, one obtains a recursive equation for the probability
distribution
\begin{equation}
\label{mft-I-pn}
P_n(x)={1\over n}\sum_{l=0}^{n-1}\sum_{y=1}^{x-1} P_{n-1}(y)\,P_l(x-y).
\end{equation}
Consequently, there are closed recursion relations for the moments.
While consistent with the exact equation (\ref{avI-rec}), the emerging
recursion relations for the higher moments are only
approximate. Analysis of these equations results in ordinary scaling
behavior, $\langle x_n^k\rangle\propto \langle x_n\rangle^k$, contrary
to Eq.~(\ref{M-grow}). Therefore, strong correlations develop,
correlations that affect the statistical characteristics.

\begin{table}
\begin{tabular}{|c|r|r|r|}
\hline
$n$& $\Gamma_n$ & $B_n$ & $G_n$ \\
\hline
3&4&3&1\\
4&8&6&2\\
5&16&13&3\\
6&30&22&6\\
7&55&39&10\\
8&98&62&18\\
9&175&117&28\\
10&310&180&50\\
11&555&367&79\\
12&986&594&144\\
13&1757&1073&249\\
14&3138&1888&432\\
15&5618&3567&756\\
\hline
\end{tabular}
\caption{The sequences $\Gamma_n$, $B_n$, and $G_n$.}
\end{table}

We now turn to model II. Here, the average $A_n$ satisfies a recursion
relation similar to Eq.~(\ref{avI-rec}),
\begin{equation}
\label{avII-rec}
A_n={2\over n}\sum_{j=0}^{n-1} A_j. 
\end{equation}  
Simplifying this equation to $A_n=\left(1+n^{-1}\right)A_{n-1}$ and
using the initial condition $A_0=1$, we obtain 
\begin{equation}
\label{avI-sol}
A_n=n+1.
\end{equation}
Numerical simulations confirm this linear growth.  Figure 3 shows that
the normalized moments remain finite asymptotically as
\begin{equation}
\label{mom-II}
\langle x_n^k\rangle \simeq \mu_k n^k.
\end{equation}
In contrast with model I, the average properly characterizes higher
order moments.  Therefore, the probability distribution $P_n(x)$
admits the scaling form
\begin{equation}
\label{phiz}
P_n(x)\simeq n^{-1}\Phi(z), \qquad z=xn^{-1},
\end{equation}
in the asymptotic limit $x,n\to\infty$ with the scaling variable
$z=xn^{-1}$ fixed (see Fig.~4).

This scaling behavior enables quantitative characterization of
extremal statistics.  Numerically, we find that the scaling function
exhibits the following extremal behaviors
\begin{equation}
\label{phi-ext}
\Phi(z)
\propto
\cases{z        &$z\to 0$,\cr 
\exp\left(-z^\kappa\right) &$z\to\infty$,\cr}
\end{equation}
with $\kappa\approx 0.4-0.5$ (see the inset to Fig.~4). The small-$z$ behavior can
be understood by considering the minimal value $x_n=2$.  This occurs
when the first element $x_0=1$ is chosen twice and therefore,
$P_n(2)=n^{-2}$.  Combining this with Eq.~(\ref{phiz}), gives
$\Phi(2n^{-1})=n^{-1}$, in agreement with the asymptotics
$\Phi(z)\propto z$ in the $z\to 0$ limit. The large-$z$ behavior is
more subtle as it depends on the entire sequence evolution. Contrary
to the small argument behavior, analysis of the maximal sequences
$x_n=2^n$ does not elucidate the large argument tail. Indeed, such
sequences occur with probability $1/n!$, much smaller than the
exponentially small probabilities dominating the large-$z$ behavior.

\begin{figure}
\centerline{\epsfxsize=7.6cm\epsfbox{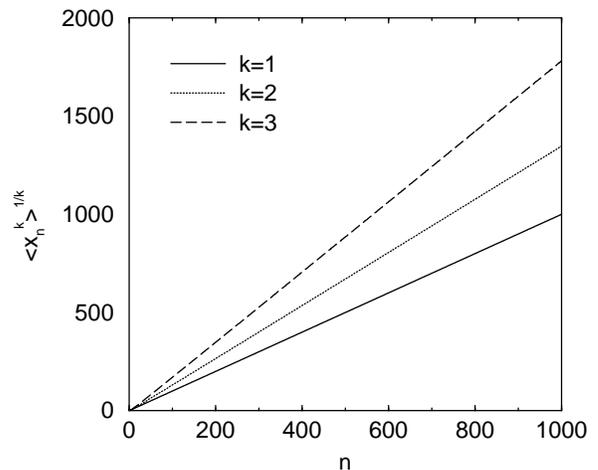}}
\caption{The moments $\langle x_n^k\rangle^{1/k}$ versus $n$ for model
II.  The data represents an average over $10^8$ realizations.}
\end{figure}

The ordinary scaling behavior indicates that mean-field theory may
provide better insight in the case of model II.  Ignoring the history
by which sequences evolve yields the following recursion for the
distribution
\begin{equation}
\label{Pnrec} 
P_n(x)=n^{-2}\sum_{l=0}^{n-1}\sum_{m=0}^{n-1}\sum_{y=1}^{x-1} P_l(y)\,P_m(x-y).
\end{equation}
Using Eq.~(\ref{phiz}), and replacing summations by integration,
equation (\ref{Pnrec}) reduces into the integral equation
\begin{equation}
\label{integeq} 
\Phi(z)=\int_0^1 {d\xi\over \xi}\int_0^1 {d\eta\over \eta}\int_0^z dz'\, 
\Phi\left({z'\over \xi}\right)
\Phi\left({z-z'\over \eta}\right). 
\end{equation}
The convolution structure suggests using the Laplace transform, and
indeed, $F(s)=\int dz e^{-sz}\Phi(z)$ satisfies a simple equation
\begin{equation}
\label{fs-eqn}
F(s)=\left[\int_0^1\,d\xi\,  F(\xi s)\right]^2.
\end{equation}
The auxiliary function $G(s)=\int_0^s ds'\, F(s')$ obeys the ordinary
differential equation $dG/ds=(G/s)^2$ from which $G(s)=s/(1+cs)$ and
then $F(s)=(1+cs)^{-2}$.  The small-$s$ behavior $F(s)=1-s$ implies
$c=1/2$.  Inverting the Laplace transform $F(s)=(1+s/2)^{-2}$ yields
the scaling function
\begin{equation}
\label{mft-phiz}
\Phi(z)=4z\,\exp\left(-2z\right).
\end{equation}
In the small-$z$ limit, mean-field theory is correct because memory is
irrelevant for $x\ll n$, (see Fig.~3). In contrast, for $x\gg n$,
memory is important and the exponent $\kappa=1$ is larger than the
numerical value $\kappa\approx 0.4-0.5$. Additionally, one may compare
the prefactors characterizing the moments defined in
Eq.~(\ref{mom-II}). Using $\mu_k=\int\, dz z^k\,\Phi(z)$ gives
$\mu_k=(k+1)!2^{-k}$ and in particular $\mu_k=1$, $3/2$, $3$, for
$k=1$, $2$, $3$. The corresponding numerical values are $1$, $1.84$,
$5.76$, respectively.  Although the history independent approximation
is quantitatively inaccurate, it still provides useful insights for
model II.

\begin{figure}
\centerline{\epsfxsize=7.6cm\epsfbox{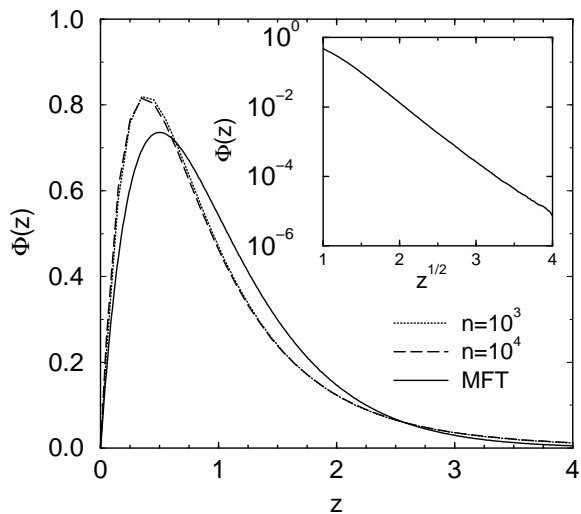}}
\caption{The scaling distribution $\Phi(z)$ vs. $z$. The distributions
were obtained from an average over $10^7$ realizations. Shown also is
the mean field theory (MFT) result (\ref{mft-phiz}). The inset shows the
large argument tail.}
\end{figure}

We have seen that the sequence growth sensitively depends on the
details of the model. In fact, the stretched exponential or algebraic
growth can be tuned by varying the recurrence rules. For example,
introducing a multiplicative factor to model II, $x_n=c(x_p+x_q)$,
leads to the algebraic growth $A_n\sim n^{2c-1}$. Functionally
different growth laws naturally emerge as well. If in model I,
$x_n=x_{n-1}+x_q$, the memory range is $0\leq q\leq [bn]$ with
$0<b<1$, one finds log-normal growth
\begin{equation}
\label{modI}
A_n\propto \exp\left(C\,\ln^2 n\right),
\end{equation}
with $C=\left[2\ln (1/b)\right]^{-1}$. This growth law is slower than
stretched exponential but faster than power law.

In summary, we have introduced a class of stochastic integer sequences
where each sequence element is a sum of two previous elements, at
least one of which is randomly chosen. While the sequence may attain a
vast range of possible values, the dynamics chooses a much narrower
range of values. Depending on the governing rules, there is a wide
spectrum of growth from algebraic to log-normal to stretched
exponential. In model I, there are infinitely many relevant scales
underlying the moments. In contrast, for model II, there is a single
scale and consequently, a mean-field approximation is qualitatively
correct.

Generally, the phase space has an intricate structure. It contains
alternating sequences of consecutive integers marking allowed and
forbidden sequence values. The gap structure consists of increasingly
complex patterns.  Interesting deterministic integer sequences such as 
the size of the phase space, the number of gaps, and the size of the
first accessible sequence underlie this phase space.

This research was supported by DOE (W-7405-ENG-36) and NSF(DMR9978902).

\end{document}